\documentclass[sigconf]{acmart}

\AtBeginDocument{
  }

\copyrightyear{2026}
\acmYear{2026}
\setcopyright{cc}
\setcctype{by}

\usepackage{amsfonts}
\usepackage{multirow}
\usepackage{stfloats}
\usepackage[english]{babel}
\usepackage{algorithm}
\usepackage{algpseudocode}

\begin{document}

\title{Variable-Length Semantic IDs for Recommender Systems}

\author{Kirill Khrylchenko}
\orcid{0009-0007-3640-8795}
\affiliation{%
  \institution{HSE University}
  \city{Moscow}
  \country{Russia}
}
\email{elightelol@gmail.com}

\renewcommand{\shortauthors}{Kirill Khrylchenko}

\begin{abstract}
Generative models are increasingly used in recommender systems, both for modeling user behavior as event sequences and for integrating large language models into recommendation pipelines. A key challenge in this setting is the extremely large cardinality of item spaces, which makes training generative models difficult and introduces a vocabulary gap between natural language and item identifiers. Semantic identifiers (semantic IDs), which represent items as sequences of low-cardinality tokens, have recently emerged as an effective solution to this problem.

However, existing approaches generate semantic identifiers of fixed length, assigning the same description length to all items. This is inefficient, misaligned with natural language, and ignores the highly skewed frequency structure of real-world catalogs, where popular items and rare long-tail items exhibit fundamentally different information requirements.
In parallel, the emergent communication literature studies how agents develop discrete communication protocols, often producing variable-length messages in which frequent concepts receive shorter descriptions. Despite the conceptual similarity, these ideas have not been systematically adopted in recommender systems.

In this work, we bridge recommender systems and emergent communication by introducing variable-length semantic identifiers for recommendation. We propose a discrete variational autoencoder with Gumbel-Softmax reparameterization that learns item representations of adaptive length under a principled probabilistic framework, avoiding the instability of REINFORCE-based training and the fixed-length constraints of prior semantic ID methods.

Our approach learns an efficient coding scheme in which popular items are represented with shorter identifiers, while long-tail and cold items receive longer, more expressive codes. Empirically, we demonstrate that the proposed method converges more stably than REINFORCE-based alternatives and enables favorable efficiency-quality trade-offs in downstream sequential recommendation, achieving comparable or improved recommendation quality under the same token budget.
These results highlight variable-length semantic identifiers as a practical and theoretically grounded representation for large-scale generative recommender systems

\end{abstract}



\maketitle

\section{Introduction}

Generative modeling has recently become a central paradigm in recommender systems. Modeling users as sequences of events and predicting the next interaction --- a fundamentally generative formulation --- has been used for many years, both to learn user representations reused across different components of recommender systems and to directly generate candidate items at the retrieval stage. Historically, however, such models were relatively small and could not realistically replace large parts of the traditional recommender stack.

This situation has changed in recent years. Following the bitter lesson\footnote{http://www.incompleteideas.net/IncIdeas/BitterLesson.html}, generative recommender models\,\cite{hstu,onerec,plum,argus} have grown significantly in scale and capacity, gradually replacing hand-crafted heuristics and classical machine learning components. Recommender systems are now following a trajectory similar to other domains where deep learning proved effective early on, including computer vision, natural language processing, and speech.

The connection between recommender systems and natural language processing goes even further. There has been growing interest in integrating recommender systems with large language models (LLMs)\,\cite{p5, plum}. The motivation is twofold: first, to improve recommendation quality by leveraging LLM capabilities such as reasoning, rich content understanding, and world knowledge; second, to enable practical conversational recommendation scenarios, allowing users to interact with recommender systems in a more flexible and transparent way.

However, integrating recommender systems with LLMs is far from straightforward. Simply extending an LLM’s vocabulary with item identifiers does not work well, as shown in prior work\,\cite{p5}. This limitation is largely due to the vocabulary gap: item identifiers correspond to high-level entities that do not align well with natural language tokens in either semantics or abstraction level. Moreover, item catalogs in real-world recommender systems are extremely large, often containing millions or even billions of items, far exceeding the typical vocabulary size of LLMs. As a result, there is a need for an intermediate representation --- a language for describing items that is both compact and compatible with natural language.

A related challenge arises when training generative models over users in recommender systems. While full softmax over the vocabulary is feasible for LLMs, computing a full softmax over a massive item catalog is prohibitively expensive. This has led to the use of approximate and often biased training techniques, such as sampled softmax with logQ correction\,\cite{logq, correctlogq}.

To address these issues, prior work\,\cite{tiger} introduced semantic identifiers (semantic IDs), which represent each item as a sequence of low-cardinality tokens. Semantic IDs are typically learned to reflect item content, making them effective for long-tail and cold-start items. They are easier to model than atomic item identifiers, allow efficient full-softmax training, and can be naturally integrated into LLM vocabularies. Moreover, while traditional recommender systems retrieve items atomically via two-tower architectures, semantic IDs enable generative retrieval: candidate items are generated by decoding entire sequences of low-cardinality tokens using beam search and then mapping these sequences back to items. This approach also offers improved control over recommendation diversity, which is difficult to achieve with standard two-tower models.

Despite these advantages, existing approaches generate semantic identifiers of fixed length, assigning the same number of tokens to every item. This design choice is misaligned with natural language, where entities are described using variable-length expressions, and where frequent or simple concepts tend to receive shorter descriptions --- a phenomenon known as Zipf’s law of abbreviation (ZLA)\,\cite{zla}. Learning variable-length semantic IDs could therefore improve efficiency by representing the same information using fewer tokens, while also bringing recommender systems closer to natural-language-based and conversational interfaces.

Interestingly, a closely related problem has been studied extensively in the field of emergent communication\,\cite{lazaridou, ec}. This line of research investigates how agents develop shared communication protocols while interacting in cooperative games. A canonical example is the Lewis signaling game\,\cite{lewis}, in which one agent observes an object and produces a symbolic message, while another agent attempts to identify the object based on that message. In effect, the agents learn to generate discrete representations --- semantic descriptions --- of objects. Importantly, prior work\,\cite{anti-efficient,lazimpa,ueda1} in emergent communication has demonstrated how to learn variable-length messages and how to encourage frequent objects to be encoded with shorter descriptions by penalizing message length during training.

At the same time, the technical approaches used in recommender systems and emergent communication differ substantially. Recommender systems typically rely on vector-quantization-based methods such as RQ-VAE\,\cite{rq-vae} or R-KMeans\,\cite{rkmeans} to construct semantic IDs, whereas emergent communication approaches are often formulated as multi-agent reinforcement learning problems and trained using REINFORCE\,\cite{reinforce, shulman}.

Inspired by emergent communication, we propose a framework for learning variable-length semantic identifiers for recommender systems. Our method produces efficient encodings in which popular items receive shorter semantic IDs, while long-tail and cold items are represented with longer, more expressive codes. This adaptive representation improves the efficiency of training and inference in generative recommender models. In addition, we bridge the gap between the recommender systems and emergent communication communities by introducing a stable and principled training approach based on discrete variational autoencoders with Gumbel-Softmax reparameterization\,\cite{gs1, gs2}, which naturally supports variable-length codes and converges more reliably than REINFORCE-based alternatives.

Our contributions are as follows:
\begin{itemize}
\item We introduce variable-length (varlen) semantic identifiers for recommender systems that efficiently allocate representational capacity by assigning shorter codes to more frequent items.
\item We demonstrate that variable-length semantic IDs provide favorable efficiency--quality trade-offs in downstream sequential recommendation tasks.
\item We propose a principled training framework based on discrete variational autoencoders with Gumbel-Softmax reparameterization, which is novel to both recommender systems and emergent communication.
\item We bridge the gap between emergent communication and recommender systems, highlighting strong conceptual and methodological connections between the two fields.
\end{itemize}

\section{Related Work}
\paragraph{Semantic IDs and Generative retrieval}
\citet{dsi} introduced \emph{generative retrieval}, where a transformer directly generates a discrete document identifier conditioned on a query, avoiding the standard retrieval pipeline based on dual encoders and approximate nearest-neighbor search.

Building on this idea, TIGER\,\cite{tiger} brought generative retrieval to recommender systems by using user interaction histories as queries and items as retrievable targets. To obtain item identifiers, TIGER proposed \emph{semantic IDs} derived from content embeddings using RQ-VAE\,\cite{rq-vae}. Beyond generative retrieval, semantic IDs were shown to improve conventional recommendation architectures: \citet{youtuberanking} replaced video-ID embeddings with semantic-ID-based embeddings in a YouTube ranking model and reported strong gains on cold-start and long-tail slices.

Semantic IDs also provide a practical interface between recommendation and large language models. \citet{plum} incorporated semantic IDs into the vocabulary of an LLM (Gemini\,\cite{gemini}) and fine-tuned it for YouTube recommendation tasks. They also proposed refinements to semantic ID construction, including a multi-resolution codebook that reduces codebook cardinality across levels, aiming to progressively encode residual information. Our work is motivated by a complementary question: whether semantic IDs must always have a fixed number of levels, or whether \emph{variable-length} identifiers can provide a better efficiency--accuracy trade-off.

There is a parallel line of work that studies sequence models for recommendation that generate over semantic IDs without explicitly using LLMs\,\cite{onerec, music_semantics, linkedin_semantic, spotify_semantic}. Using semantic IDs as tokens enables large-vocabulary generation and can provide more expressive matching mechanisms compared to dot-product retrieval in two-tower models.

\paragraph{Learning discrete representations}
Discrete representations of objects are commonly learned using two families of variational autoencoder (VAE)\,\cite{vae} approaches. The first family consists of discrete VAEs (dVAEs) based on the Gumbel-Softmax reparameterization\,\cite{gs1, gs2}, which enables approximately differentiable sampling from categorical distributions. A prominent example is DALLE-1\,\cite{dalle1}, where a Gumbel-Softmax dVAE is first trained to encode images into discrete latent tokens, and a transformer is subsequently trained to predict these tokens from paired textual captions.

The second family is based on vector quantization\,\cite{vq}, where gradients are propagated from selected codebook entries to the encoder using straight-through estimation\,\cite{ste}. In this setting, the explicit variational formulation with KL-divergence and priors is discarded, making it non-trivial to incorporate structured priors such as a length prior for penalizing message length.
In recommender systems, early work predominantly relied on RQ-VAE\,\cite{rq-vae}, a hierarchical extension of VQ-VAE\,\cite{vq-vae} that enforces a residual encoder structure: codewords are selected sequentially, each encoding the residual left after subtracting previously selected codewords. More recent studies\,\cite{onerec, snapchat} demonstrate that simpler methods such as Residual KMeans (R-KMeans)\,\cite{rkmeans}, which iteratively applies the KMeans algorithm\,\cite{kmeans} to residual vectors, can outperform RQ-VAE while being easier to train and scale.

Notably, all of the above approaches produce semantic IDs of a fixed length that is shared across all objects. To the best of our knowledge, existing VQ- or Gumbel-Softmax-based methods have not been adapted to learn variable-length semantic identifiers.

\paragraph{Emergent communication and variable-length messages}
Emergent communication (EC)\,\cite{lazaridou,ec} studies forms of communication that arise between agents through interaction. 
One of the most common experimental setups in this field is the cooperative Lewis game\,\cite{lewis}, which introduces two agents: a sender that observes an object and communicates a symbolic message, and a receiver that attempts to identify or reconstruct the object based on this message. 
Several variants of this game have been studied, including referential games, where the receiver must select the correct object among distractors, and reconstruction games, where the receiver is tasked with reconstructing the object itself. 
EC problems are typically framed as multi-agent reinforcement learning tasks, which has led to the widespread use of REINFORCE\,\cite{reinforce}.

Early works in EC already explored variable-length messages\,\cite{Titov}. \citet{anti-efficient} observed that, unlike natural language where frequently occurring concepts tend to have shorter descriptions (Zipf's law of abbreviation\,\cite{zla}), unconstrained training with variable-length messages often leads to \emph{anti-efficient coding}, in which frequent objects receive longer descriptions. 
A simple and effective remedy is to introduce an explicit penalty on message length\,\cite{anti-efficient, lazimpa, ueda1}. 
Follow-up work also explored improving communication efficiency by injecting noise into the communication channel\,\cite{ueda1} or by making the receiver \emph{impatient}\,\cite{lazimpa} by averaging the reconstruction loss over all prefixes of the message.

There is a direct analogy between Lewis games and VAEs: the sender corresponds to the encoder, the receiver to the decoder, and the reconstruction objective naturally induces a reconstruction loss.
Although the Gumbel-Softmax relaxation has been explored in emergent communication settings\,\cite{Titov, anti-efficient}, \citet{anti-efficient} concluded that REINFORCE provides more stable training for discrete messages. However, their analysis does not employ a full variational formulation with an explicit KL-divergence term.

More recently, \citet{ueda2} showed that a Lewis game can be formulated as a discrete VAE with an explicit prior over message lengths. 
They demonstrated that a energy-based prior over message lengths naturally induces a message length penalty and derived the corresponding ELBO objective, but still optimized it using REINFORCE rather than the Gumbel-Softmax reparameterization.

Finally, it is important to note that most EC studies operate at relatively small scales, with limited vocabularies (e.g., tens of symbols) and small datasets (e.g., thousands of unique objects).

To summarize, in recommender systems, semantic IDs have primarily been studied in fixed-length form, despite the highly non-uniform frequency structure of real-world catalogs. 
In contrast, emergent communication research has extensively explored variable-length messages and efficiency-inducing regularization, but mostly at small scales and without fully adopting a variational training framework. 
As a result, it remains unclear how to learn variable-length semantic identifiers that are both efficient and scalable to large-scale recommendation problems.

In this work, we address this gap by formulating variable-length semantic ID learning as a discrete VAE problem with an explicit probabilistic model of code length. 
Our approach enables learning variable-length semantic IDs at scale and naturally connects ideas from emergent communication with modern recommender system architectures, while further developing them within a scalable variational framework.
\section{Method}
\subsection{Problem Setup and Notation}
Let $\mathcal{I}$ denote a finite set of items, where each item is associated with an embedding $x \in \mathbb{R}^d$. 
Our goal is to represent each item by a discrete code sequence $z = (z_1, \dots, z_L)$, where $L \leq T$ denotes the code length and each symbol $z_t \in \mathcal{V}$ is drawn from a shared vocabulary.
In the terminology of emergent communication, the sequence $z$ can be viewed as a \emph{message} describing the item.

In many existing recommender-system approaches to discrete item representations, different positions in the code are treated as semantically independent: each level has its own vocabulary and its own embedding space, and tokens at different positions are modeled separately. 
Such representations are convenient from an engineering perspective, but they do not naturally correspond to a single shared symbolic system.

In contrast, similarly to emergent communication, we assume that all messages are drawn from a single shared language, meaning that the same vocabulary is used across all positions in the sequence.
Under this assumption, a symbol carries the same semantic meaning regardless of its position in the message.

\subsection{Generative Model}
For an item embedding $x \in \mathbb{R}^d$, a latent discrete message $z$ and its length $L \leq T$, we assume the following generative model:
\begin{equation*}
p(x, z, L) = p(L)\, p(z \mid L)\, p(x \mid z_{1:L}).
\end{equation*}
We represent $z=(z_1,\dots,z_T)$ with padding, where tokens after the stopping time $L$ are deterministically equal to \texttt{pad}. The decoder depends only on the prefix $z_{1:L}.$

\paragraph{Prior over symbols.}
We choose a position-independent prior over message content:
\begin{equation*}
p(z \mid L)
=
\prod_{t=1}^L p_{\mathcal{V}}(z_t)
\prod_{t=L+1}^T \mathbb{I}\{z_t=\texttt{pad}\},
\qquad
p_{\mathcal V}(z_t)=\frac{1}{|\mathcal V|}.
\end{equation*}
Such a choice promotes exploration and encourages the model to utilize the shared vocabulary more uniformly, rather than collapsing to a small subset of symbols.

\paragraph{Prior over length.}
We use a truncated geometric distribution as a length prior:
\begin{equation*}
p(L)=\frac{(1-\alpha)^{L-1}\alpha}{1-(1-\alpha)^T},
\end{equation*}
where $\alpha\in(0,1)$ is the per-step stopping probability.
In the variational objective, this form translates into an explicit message length penalty and is equivalent up to reparameterization to the energy-based length prior of \cite{ueda2}.

\subsection{Variational Inference}
For the generative model defined above, the true posterior $p(z, L \mid x)$ is intractable.
As in standard variational autoencoder formulations, we therefore introduce a parameterized approximation
$q(z, L \mid x)$, which serves as the encoder that maps an item embedding $x$ to a discrete message $z$ and its length $L$.

We assume the following form of the approximate posterior:
\begin{equation*}
q(z, L \mid x) =
q(L \mid x, z_{1:L})\;
\prod_{t=1}^{L} q(z_t \mid z_{1:t-1}, x)
\prod_{t=L+1}^{T} \mathbb{I}\{z_t = \texttt{pad}\}\;
.
\end{equation*}
Under this parameterization, symbols are generated autoregressively, and the probability of a particular length $L$
depends only on the prefix $z_{1:L}$ rather than on the full message.

This choice reflects how we would like message length to be determined in practice.
Deciding when to stop transmitting a message --- or, equivalently, when an object has been described sufficiently --- should be a sequential decision that depends both on the object itself and on the content already communicated.

Accordingly, we model message termination via per-step stopping probabilities.
At each step $l$, the encoder predicts the probability of stopping after emitting the $l$-th symbol,
\[
q(s_l = 1 \mid x, z_{1:l}),
\]
where $s_l$ denotes the stopping event.
These probabilities induce the following distribution over message lengths:
\begin{align*}
q(L = l \mid x, z_{1:l})
=
q(s_l = 1 \mid x, z_{1:l})
\prod_{t=1}^{l-1} \bigl(1 - q(s_t = 1 \mid x, z_{1:t})\bigr).
\end{align*}

We use the stopping probabilities as a convenient parameterization of the length distribution,
but formulate all objectives and derivations in terms of the random variable $L$.
This is possible because under our construction the length posterior satisfies
$q(L \mid x, z) = q(L \mid x, z_{1:L})$, which allows us to work directly with $L$ while retaining the sequential stopping interpretation.

\paragraph{Length modeling without an EOS symbol.}
Unlike standard emergent communication approach, we do not introduce an explicit end-of-sequence (EOS) symbol into the vocabulary.
Instead, message length is modeled separately via stopping probabilities and the latent variable $L$.
As a result, the expected message length is controlled solely by the length model and its prior, rather than implicitly depending on the vocabulary size or symbol-level probabilities.

\subsubsection{ELBO}
The model is trained by maximizing the evidence lower bound (ELBO), which for our generative model takes the standard form:
\begin{equation*}
\mathrm{ELBO}
=
\mathbb{E}_{q(z, L \mid x)}\!\left[ \log p(x \mid z_{1:L}) \right]
-
\mathrm{KL}\!\left( q(z, L \mid x)\,\|\, p(z, L) \right).
\end{equation*}
The detailed derivation of the final objective is provided in Appendix~\ref{app:elbo_derivation}.
Below, we present the resulting training loss in a form that directly corresponds to what is optimized in practice.

\paragraph{Final objective.}
For a given training example, the ELBO can be written as a sum of three terms:
\begin{equation*}
\mathcal{L}
=
\mathcal{L}_{\mathrm{recon}}
+
\mathcal{L}_{\mathrm{reg}}^{\mathrm{vocab}}
+
\mathcal{L}_{\mathrm{reg}}^{\mathrm{length}}.
\end{equation*}

\paragraph{Reconstruction loss.}
The reconstruction term aggregates reconstruction errors over all message prefixes,
weighted by their probabilities:
\begin{equation*}
\mathcal{L}_{\mathrm{recon}}
=
-
\mathbb{E}_{q(z \mid x)}
\left[
\sum_{l=1}^{T}
q(L=l \mid x, z_{1:l})
\log p(x \mid z_{1:l})
\right].
\end{equation*}
In practice, to stabilize training, we mix the length distribution with a uniform distribution over prefixes,
using weights $0.9$ and $0.1$, respectively.

\paragraph{Vocabulary regularization.}
The vocabulary regularization term takes the following form:
\begin{equation*}
\mathcal{L}_{\mathrm{reg}}^{\mathrm{vocab}}
=
\mathbb{E}_{q(z \mid x)}
\left[
\sum_{l=1}^{T}
q(L \geq l \mid x, z_{1:l})\,
\mathrm{KL}\!\left(
q(z_l \mid x, z_{1:l-1})
\,\Vert\,
p_{\mathcal V}
\right)
\right].
\end{equation*}
where
\(
q(L \geq l \mid x, z_{1:l})
\)
is the probability that the $l$-th position is included in the message.
This term regularizes symbol predictions and encourages broad usage of the shared vocabulary.

\paragraph{Length regularization.}
The length regularization term is given by:
\begin{equation*}
\mathcal{L}_{\mathrm{reg}}^{\mathrm{length}}
=
\mathbb{E}_{q(z \mid x)}
\left[
\lambda\, \mathbb{E}_{q(L \mid x, z)}[L]
-
H\!\left(q(L \mid x, z)\right)
\right].
\end{equation*}
The expected-length term is a direct penalty on message length, analogous to length penalties commonly used in emergent communication to encourage efficient codes.
The entropy term prevents premature collapse to a single length and promotes diversity of message lengths.

For clarity, we provide a concise pseudocode description of the full training objective computation in Algorithm~\ref{alg:varlen_dvae_loss} of Appendix~\ref{app:elbo_derivation}.

\paragraph{Discussion}
These terms interact in a non-trivial way.
Increasing the message length generally improves reconstruction quality, but also increases both vocabulary and length regularization costs.
Conversely, shorter messages reduce regularization but limit the amount of information that can be conveyed.
The final representation therefore emerges from a balance between reconstruction accuracy, vocabulary usage, and message length.

\subsection{Residual Encoding with Soft Relaxation}\label{sec:soft}
In classical residual quantization, each step selects a single codebook vector via a hard assignment and subtracts it from the residual representation. In contrast, our encoder uses a Gumbel-Softmax relaxation to produce a soft assignment over the vocabulary, which corresponds to taking the expected codebook embedding under this distribution. This expected embedding is subtracted from the hidden state, yielding a differentiable analogue of residual quantization. Algorithm~\ref{alg:varlen_dvae_loss} in Appendix~\ref{app:elbo_derivation} provides a detailed pseudocode description of this procedure.

\subsection{Optimization}
The proposed model is trained end-to-end with standard backpropagation using the Gumbel-Softmax relaxation for discrete latent variables.

To improve training stability, we adopt a $\beta$-VAE scheme\,\cite{betavae}, gradually increasing the weight of the KL regularization from zero during an initial warm-up period to prevent posterior collapse.

We further apply the free bits technique\,\cite{freebits}, which enforces a minimum information budget per latent step.

\section{Experiments}
We aim to answer the following research questions:

\begin{itemize}
\item \textbf{RQ1:} Can we learn efficient varlen semantic IDs?

\item \textbf{RQ2:} Do varlen semantic IDs work well for downstream recommendation tasks?

\item \textbf{RQ3:} Is our dVAE-based approach a viable alternative to REINFORCE-based emergent communication methods?

\item \textbf{RQ4:} How scalable are varlen semantic IDs?
\end{itemize}

\subsection{Experimental Setup}
In this section, we describe our datasets, baselines, evaluation protocol, and implementation details.

\paragraph{Datasets}

\begin{table}[t]
\centering
\small
\setlength{\tabcolsep}{6pt}
\caption{Dataset statistics after preprocessing. Head share denotes the fraction of interactions attributed to the top-30k most popular items.}
\label{tab:datasets}
\begin{tabular}{lrrrr}
\toprule
\textbf{Dataset} & \textbf{\#Items} & \textbf{\#Events} & \textbf{Emb. dim} & \textbf{Head share} \\
\midrule
Yambda & 268K & 82M & 128 & 0.747 \\
VK-LSVD & 900K & 95M & 64 & 0.276 \\
Amazon T\&G & 305K & 12M & 128 & 0.597 \\
\bottomrule
\end{tabular}
\end{table}

We evaluate the proposed method on three large-scale recommendation datasets: Yambda, VK-LSVD, and Amazon Toys \& Games. Table~\ref{tab:datasets} summarizes their key statistics.

\textbf{Yambda}\,\cite{yambda} is a large-scale music recommendation dataset containing billions of user-item interactions.
We keep likes only and filter out items with fewer than 16 training-period interactions. 
We use the precomputed item embeddings provided with the dataset.

\textbf{VK-LSVD}\,\cite{vklsvd} is a large-scale short-video recommendation dataset.
We use a version containing 10\% of users, retain only likes, and apply the same minimum-interaction threshold as above.
Item embeddings are provided as part of the dataset.

\textbf{Amazon Toys \& Games}\,\cite{amazondataset} contains product reviews in the Toys and Games category.
We treat ratings $\geq 4$ as positive feedback and discard the remaining interactions, filtering out items with fewer than 5 positive interactions.
Item embeddings are computed from textual product metadata using a pretrained text embedding model Embedding Gemma\,\cite{gemma}.

\subsubsection{Baselines}

To validate our approach against established methods, we consider several baselines.

\textbf{R-KMeans}\,\cite{rkmeans} is a widely used semantic ID construction method in recommender systems.
It iteratively applies the KMeans algorithm\,\cite{kmeans}: first to the original item embeddings, and then to successive residuals obtained by subtracting previously assigned centroids.
At each iteration, the index of the selected centroid becomes a token in the semantic ID.
R-KMeans is trained solely on the item catalog and does not utilize user-item interaction data.

We also implement a \textbf{REINFORCE-based approach} commonly used in emergent communication research\,\cite{egg}.
The sender (encoder) is optimized using policy gradients, while the receiver (decoder) is trained via standard backpropagation\,\cite{shulman}.
The objective consists of a reconstruction loss, an entropy regularizer for the sender, and, in the variable-length setting, an explicit message length penalty.
Variable-length messages are obtained by introducing an explicit EOS symbol into the vocabulary.
For a fair comparison, we use the same encoder and decoder architectures as in our dVAE model.
We found that REINFORCE-based training with single-layer LSTM encoders, commonly used in emergent communication\,\cite{egg}, was unstable at the scale considered in our experiments.

\subsubsection{Evaluation Metrics}\label{sec:eval}
We adopt a temporal train--test split.
For Yambda and VK-LSVD, the last week of interactions is used for testing, while for Amazon Toys \& Games we use the last four weeks.
This split is used both for downstream recommendation evaluation and for analyzing semantic ID properties on cold items.
In particular, evaluating semantic ID length and reconstruction quality on cold items allows us to assess the generalization ability of learned identifiers beyond the training catalog.

\paragraph{Semantic ID evaluation}
We evaluate semantic IDs along three dimensions: reconstruction quality, vocabulary utilization, and coding efficiency.

\emph{Reconstruction quality} is measured on the final generated message for each item.
Specifically, for each item we generate a single semantic ID with an inferred length $L$ and compute the reconstruction loss using the prefix $z_{1:L}$.
Reconstruction errors are then averaged under the empirical unigram distribution of interactions, reflecting the expected reconstruction quality on data rather than uniform averaging over the catalog.

\emph{Vocabulary utilization} is assessed using \emph{micro-averaged token perplexity}.
We compute the empirical distribution of all generated semantic tokens across all positions and items, and report perplexity as
\[
\mathrm{PPL} = \exp\!\left( - \sum_{v \in V} p(v)\, \log p(v) \right),
\]
where $p(v)$ denotes the empirical frequency of token $v$.
Since our model uses a single shared vocabulary across positions, micro perplexity provides a natural measure of how effectively the discrete capacity is utilized.
We do not report codebook usage separately, as it is effectively $100\%$ for both dVAE-based models and R-KMeans in all our experiments and therefore not informative.
For additional analysis, we report position-wise (per-step) token perplexity in Appendix~\ref{app:exp}.

\emph{Coding efficiency} is evaluated by analyzing the relationship between semantic ID length and item popularity.
We compute semantic IDs for all items and group them by code length.
For each length bucket, we report the mean and maximum item popularity (measured as the number of user-item interactions).
We further compare average semantic ID length under a uniform item distribution and under the empirical unigram distribution of interactions.

\paragraph{Downstream sequential recommendation}
To evaluate the usefulness of semantic IDs in downstream tasks, we consider a next-item prediction setting with a transformer-based sequential recommender.
Our goal is not to achieve state-of-the-art performance, but to compare fixed-length and varlen semantic representations under controlled conditions.

We report Recall@K to measure ranking quality and Coverage@K to assess recommendation diversity.
Following \citet{time_to_split}, we sample one test interaction per user from the test period and use the preceding interaction history as input.
To reduce evaluation cost, we randomly subsample test users, retaining 5\% of users for Yambda and 10\% for VK-LSVD.
All models are trained and evaluated under a fixed token budget (512 tokens) for the user history.
We additionally report the average number of user-item events that fit into this budget, which directly reflects the efficiency gains enabled by shorter semantic IDs.

\subsubsection{Implementation Details}

\paragraph{Encoder and decoder}
The encoder architecture follows the residual formulation introduced in Section \ref{sec:soft}, where semantic IDs are constructed autoregressively using a shared vocabulary and a differentiable relaxation of residual quantization.

The decoder is implemented as a causal transformer\,\cite{transformer} that reconstructs item embeddings from semantic ID prefixes.
Because the decoder is autoregressive, a single forward pass over the full message produces reconstructions conditioned on all prefixes, enabling reconstruction losses to be computed for every prefix length.

Reconstruction loss is defined as mean squared error between $\ell_2$-normalized embeddings, which is equivalent to cosine distance.

\paragraph{Transformer backbone}
All transformer-based components in our models follow the \texttt{nanoGPT speedrun} implementation\,\cite{nanogpt_speedrun},
including YARN positional encodings\,\cite{yarn}, ReLU-squared activations\,\cite{relusq}, and pre-normalization with parameter-free RMSNorm\,\cite{rmsnorm}.

\paragraph{Tuning dVAE}
We tune four key hyperparameters of the dVAE model:

\begin{itemize}
    \item \textbf{Gumbel-Softmax temperature.}
    The temperature is annealed from $\tau=1$ to $\tau_{\min}$.
    We use $\tau_{\min}=0.5$ for Yambda and $\tau_{\min}=0.7$ for VK-LSVD and Amazon.

    \item \textbf{$\beta$-annealing.}
    The KL weight is increased from 0 to $\beta_{\max}=0.002$ during the first training epoch using cosine schedule.

    \item \textbf{Free bits.}
    A free-bits threshold of 2.0 improves performance on Yambda, while having negligible effect on VK-LSVD and Amazon.

    \item \textbf{Length cost $\lambda$.}
    Larger values encourage shorter messages, while smaller values allow longer codes.
    Exploring $\lambda \in [0,12]$ was sufficient to obtain the desired length--quality trade-off.
\end{itemize}

\paragraph{Tuning REINFORCE}
The sender is regularized with an entropy term, whose weight is initialized to $0.03$ and annealed to $10^{-3}$ over the first 6{,}000 training steps (approximately half an epoch on Yambda).
In the variable-length setting, the message length penalty is annealed from 0 to 0.02 over the same interval.
To reduce gradient variance, we use separate running-mean baselines for the reconstruction loss, entropy regularization, and length penalty.
All remaining hyperparameters (batch size, optimizer, architecture, vocabulary size, and maximum length) are kept identical to those used for the dVAE model.

\paragraph{Training setup}
We use a batch size of 8192 and optimize all models using AdamW with learning rate $10^{-3}$ and weight decay $10^{-5}$.
Training is performed for 5 epochs on Yambda and VK-LSVD, and for 10 epochs on Amazon Toys \& Games.
The maximum semantic ID length is set to 5, and the vocabulary size is fixed to 4096.

\paragraph{Sequential recommender}
For downstream evaluation, we use a decoder-only transformer trained autoregressively over sequences of semantic IDs.
To resolve collisions between semantic IDs, we append a special unique identifier token to each semantic ID, following prior work\,\cite{tiger}.

The sequential recommender uses 8 transformer layers with hidden size 512.
Transformer layers are optimized using the NorMuon optimizer\,\cite{normuon, muon}, while embeddings and output heads are trained with AdamW.
Training follows a nanochat-style pretraining regime\,\cite{nanochat}, where sequences are formed by reshaping batches of tokens.
We use gradient accumulation to achieve an effective batch size of 262{,}144 tokens.

We provide the full implementation, including all configs and hyperparameters for every experiment, in the accompanying repository\footnote{\url{https://github.com/KhrylchenkoKirill/varlen_semantic_ids}} to ensure reproducibility.

\subsection{RQ1: Can we learn efficient varlen semantic IDs?}

\begin{table*}[t]
\centering
\small
\setlength{\tabcolsep}{4pt}
\caption{
Semantic ID reconstruction quality and efficiency. $\mathbb{E}_{p_{\text{data}}}[L]$ denotes average semantic ID length under $p_{\text{data}}$ distribution.
}
\begin{tabular}{lcccccccccccc}
\toprule
& \multicolumn{4}{c}{\textbf{Yambda}} 
& \multicolumn{4}{c}{\textbf{VK-LSVD}} 
& \multicolumn{4}{c}{\textbf{Amazon Toys\&Games}} \\
\cmidrule(lr){2-5} \cmidrule(lr){6-9} \cmidrule(lr){10-13}
\textbf{Method}
&  $\mathcal{L}_{\text{recon}}$ & $\mathcal{L}_{\text{recon}}^{\text{cold}}$ & $\mathbb{E}_{p_{\text{data}}}[L]$ & 
PPL
&  $\mathcal{L}_{\text{recon}}$ & $\mathcal{L}_{\text{recon}}^{\text{cold}}$ & $\mathbb{E}_{p_{\text{data}}}[L]$ & 
PPL
&  $\mathcal{L}_{\text{recon}}$ & $\mathcal{L}_{\text{recon}}^{\text{cold}}$ & $\mathbb{E}_{p_{\text{data}}}[L]$ & 
PPL \\
\midrule
R-KMeans
& 0.038 & 0.063 & 5 & 3721 
& 0.102 & 0.149 & 5 & 4010 
& 0.094 & 0.138 & 5 & 3288 \\
dVAE (fixed-length)
& 0.025 & 0.054 & 5 & 3557 
& 0.123 & 0.147 & 5 & 3959 
& 0.076 & 0.170 & 5 & 3797 \\
dVAE (varlen, $\lambda{=}\lambda_1$)
& 0.025 & 0.057 & 3.62 & 3574 
& 0.146 & 0.160 & 4.06 & 3913 
& 0.068 & 0.167 & 4.63 & 3800\\
dVAE (varlen, $\lambda{=}\lambda_2$)
& 0.026 & 0.061 & 2.96 & 3609 
& 0.155 & 0.176 & 3.45 & 3879 
& 0.071 & 0.167 & 3.01 & 3623  \\
dVAE (varlen, $\lambda{=}\lambda_3$)
& 0.029 & 0.076 & 2.29 & 3564 
& 0.173 & 0.193 & 2.94 & 3839
& 0.076 & 0.182 & 2.38  & 3339  \\
\bottomrule
\end{tabular}
\label{tab:rq1_quality}
\end{table*}

\begin{table*}[t]
\centering
\small
\setlength{\tabcolsep}{4pt}
\caption{
Item popularity statistics by semantic ID length on the training data
}

\begin{tabular}{c ccc ccc ccc}
\toprule
& \multicolumn{3}{c}{\textbf{Yambda}} 
& \multicolumn{3}{c}{\textbf{VK-LSVD}} 
& \multicolumn{3}{c}{\textbf{Amazon Toys\&Games}} \\
\cmidrule(lr){2-4} \cmidrule(lr){5-7} \cmidrule(lr){8-10}
\textbf{Semantic ID length}  
& Mean Pop. & Max Pop. & \# Items
& Mean Pop. & Max Pop. & \# Items
& Mean Pop. & Max Pop. & \# Items \\
\midrule
1 & 2248.9 & 82097 & 7562   & 165.2 & 12319 & 18572  & 392.0 & 18275 & 5020   \\
2 & 1154.1 & 23034 & 12485  & 151.3 & 6222  & 59305  & 99.8  & 2705  & 31384  \\
3 & 292.2  & 10082 & 28421  & 115.0 & 3600  & 94966  & 39.2  & 724   & 31412  \\
4 & 214.1  & 10037 & 80505  & 104.8 & 3079  & 159327 & 28.7  & 632   & 54192  \\
5 & 125.4  & 7027  & 112276 & 89.4  & 2695  & 477542 & 18.6  & 575   & 152429 \\
\bottomrule
\end{tabular}
\label{tab:rq1_pop_by_length}
\end{table*}

\begin{table}[t]
\centering
\small
\setlength{\tabcolsep}{6pt}
\caption{
Average semantic ID length under catalog and data distributions
}

\begin{tabular}{lcccc}
\toprule
& \multicolumn{2}{c}{\textbf{Train}} 
& \multicolumn{2}{c}{\textbf{Cold}} \\
\cmidrule(lr){2-3} \cmidrule(lr){4-5}
\textbf{Dataset}
& $\mathbb{E}_{p_{\text{catalog}}}[L]$
& $\mathbb{E}_{p_{\text{data}}}[L]$
&  $\mathbb{E}_{p_{\text{catalog}}}[L]$
& $\mathbb{E}_{p_{\text{data}}}[L]$ \\
\midrule
Yambda      
& 4.15 & 2.96 
& 4.53 & 4.40 \\
VK-LSVD     
& 3.66 & 3.45 
& 4.16 & 3.94 \\
Amazon T\&G 
& 4.16 & 3.01 
& 4.49 & 4.49 \\
\bottomrule
\end{tabular}
\label{tab:rq1_len_train_cold}
\end{table}

In this section, we study whether variable-length semantic IDs can achieve \emph{efficient coding}.
By efficiency, we mean two properties:
(1) semantic reconstruction quality comparable to fixed-length representations (fixed-length semantic IDs), despite using fewer tokens on average, and
(2) adaptation of code length to item popularity, such that frequent items receive shorter descriptions, consistent with the Zipfian law of abbreviation.

We first verify that our variational formulation is competitive with established semantic ID methods.
As shown in Table~\ref{tab:rq1_quality}, fixed-length dVAE achieves reconstruction quality comparable to or better than R-KMeans on Yambda and Amazon Toys\&Games, while slightly underperforming on VK-LSVD.
This confirms that the proposed dVAE formulation is a viable alternative to residual quantization baselines.

\paragraph{Quality--efficiency trade-off.}
Table~\ref{tab:rq1_quality} shows that varlen dVAE preserves reconstruction quality close to fixed-length baselines while substantially reducing the average number of tokens per item.
Increasing the length cost $\lambda$ leads to progressively shorter codes, with only a modest degradation in reconstruction loss.
This demonstrates that allowing variable-length descriptions enables a favorable efficiency--quality trade-off.

\paragraph{Popularity--length relationship.}
Table~\ref{tab:rq1_pop_by_length} shows that learned semantic ID lengths strongly correlate with item popularity.
Across all datasets, items assigned shorter codes have significantly higher average and maximum popularity, while longer codes are predominantly used for infrequent items.
This behavior emerges without explicitly encoding popularity information and closely matches the principle of efficient communication, where frequent items receive shorter descriptions.

\paragraph{Uniform vs. data-weighted statistics.}
Table~\ref{tab:rq1_len_train_cold} compares average semantic ID length under catalog-level averaging ($p_{\text{catalog}}$) and data-weighted averaging based on the empirical interaction distribution ($p_{\text{data}}$).
On the training split, $p_{\text{data}}$-weighted averages are consistently shorter, reflecting the dominance of frequent items, which are assigned more compact codes.
In contrast, cold items exhibit substantially longer semantic IDs, approaching the $p_{\text{catalog}}$ baseline. This behavior is intuitive: when an item is unfamiliar to the sender and lacks sufficient interaction signal, it is encoded using a longer and more descriptive message.

Overall, RQ1 shows that varlen semantic IDs enable efficient semantic coding: reconstruction quality is largely preserved while code length adapts to item frequency, allocating shorter representations to popular items and longer ones to rare and cold items.

\subsection{RQ2: Do varlen semantic IDs work well for downstream recommendation tasks?}

\begin{table*}[t]
\centering
\small
\setlength{\tabcolsep}{5pt}
\caption{
Sequential recommendation performance under a fixed token budget.
Tokens and Events denote the number of historical tokens and interaction events for the test users. $L_{\text{cand}}$ denotes the average length of semantic IDs generated for candidate items during evaluation
}
\begin{tabular}{lccccc ccccc}
\toprule
\textbf{Method}
& \multicolumn{5}{c}{\textbf{Yambda}}
& \multicolumn{5}{c}{\textbf{VK-LSVD}} \\
\cmidrule(lr){2-6} \cmidrule(lr){7-11}
& Recall@100 & Tokens & Events & $L_{\text{cand}}$ & Cov@100
& Recall@100 & Tokens & Events & $L_{\text{cand}}$ & Cov@100 \\
\midrule
R-KMeans
& 0.1637 \, (+0.0\%) & 364.8 & 61.49 & 6.00 & 18.2 \, (+0.0\%)
& 0.0363 \, (+0.0\%) & 326.7 & 55.87 & 6.00 & 5.1 \, (+0.0\%) \\

dVAE (fixed-length)
& 0.1645 \, (+0.5\%) & 364.8 & 61.49 & 6.00 & 20.8 \, (+14.3\%)
& 0.0410 \, (+12.9\%) & 326.7 & 55.87 & 6.00 & 6.6 \, (+29.4\%) \\

dVAE (varlen, $\lambda{=}1$)
& 0.1676 \, (+2.4\%) & 331.2 & 71.72 & 4.47 & 20.9 \, (+14.8\%)
& 0.0415 \, (+14.3\%) & 308.5 & 66.60 & 4.25 & 8.2 \, (+60.8\%) \\

dVAE (varlen, $\lambda{=}2$)
& 0.1794 \, (+9.6\%) & 287.7 & 88.93 & 3.05 & 25.1 \, (+37.9\%)
& 0.0418 \, (+15.2\%) & 302.6 & 70.49 & 3.96 & 9.0 \, (+76.5\%) \\

dVAE (varlen, $\lambda{=}3$)
& \textbf{0.1820} \, (+11.2\%) & \textbf{277.8} & \textbf{93.13} & \textbf{2.86} & 
\textbf{25.8} \, (+41.8\%)
& \textbf{0.0459} \, (+26.5\%) & \textbf{291.7} & \textbf{78.24} & \textbf{3.50} & \textbf{10.3} \, (+102.0\%) \\
\bottomrule
\end{tabular}
\label{tab:rq2_seqrec}
\end{table*}

In this section, we evaluate varlen semantic IDs in a downstream sequential recommendation setting.
Our goal is not to outperform specialized state-of-the-art recommenders, but to assess how varlen semantic IDs compare to fixed-length representations under a controlled and fixed evaluation setup.
Specifically, we fix the recommender architecture, training procedure, and token budget, and vary only the method used to construct semantic IDs.

\paragraph{Recommendation quality.}
Table~\ref{tab:rq2_seqrec} shows that fixed-length dVAE, trained on user-item interaction data, outperforms R-KMeans in terms of Recall@100 on both Yambda and VK-LSVD.
This indicates that learning semantic IDs through a variational objective grounded in interaction data yields higher-quality representations for downstream recommendation.

On both Yambda and VK-LSVD, varlen dVAE consistently outperforms its fixed-length counterpart in terms of recall.
Overall, allowing variable-length semantic IDs does not degrade downstream recommendation quality and can provide measurable gains in practice.

\paragraph{Efficiency under a fixed token budget.}
By producing shorter semantic IDs, varlen models allow substantially more user-item events to be represented within the same token budget.

\paragraph{Coverage.}
As shown in Table~\ref{tab:rq2_seqrec}, varlen semantic IDs achieve higher Coverage@100, indicating a broader set of recommended items.
This suggests that more efficient allocation of representational capacity leads to increased recommendation diversity.

Taken together, these results show that varlen semantic IDs are effective for downstream sequential recommendation: they improve recall, increase efficiency under fixed token budgets, and improve recommendation coverage.

\subsection{RQ3: Is our dVAE-based approach a viable alternative to REINFORCE-based emergent communication methods?}

\begin{table}[t]
\centering
\small
\setlength{\tabcolsep}{6pt}
\caption{
Comparison of dVAE and REINFORCE-based training on Yambda
}
\begin{tabular}{lccccc}
\toprule
\textbf{Method}
& $\mathcal{L}_{\text{recon}}$
& $\mathcal{L}_{\text{recon}}^{\text{cold}}$
& $\mathbb{E}_{p_{\text{data}}}[L]$
& PPL \\
\midrule
dVAE (fixed-length)
& 0.0246
& 0.0536
& 5.00
& 3558 \\

dVAE (varlen)
& 0.0294
& 0.0763
& 2.29
& 3564\\

REINFORCE (fixed-length)
& 0.0259
& 0.0607
& 5.00
& 2720\\

REINFORCE (varlen)
& 0.0528
& 0.1499
& 2.65
& 316\\
\bottomrule
\end{tabular}
\label{tab:rq3_reinforce}
\end{table}
We compare Gumbel-Softmax dVAE training with a REINFORCE-based approach for learning semantic IDs on Yambda.

\paragraph{Fixed-length setting.}
As shown in Table~\ref{tab:rq3_reinforce}, fixed-length dVAE consistently achieves lower reconstruction error and better codebook utilization than REINFORCE, both on the full item set and on cold items.
This indicates that, even when message length is fixed, the variational formulation provides a more effective optimization objective.

\paragraph{Variable-length setting.}
In the variable-length regime, REINFORCE-based training becomes less stable.
We experimented with different length penalty schedules and replaced the residual encoder with a one-layer LSTM identical to popular EGG\,\cite{egg} implementation, but observed consistently degraded reconstruction quality and a collapse in codebook perplexity.
This suggests that, while REINFORCE can be made to work, it is harder to make it work at scale.

Overall, these results indicate that dVAE provides a more stable and practical alternative to REINFORCE for learning variable-length semantic IDs at scale.

\subsection{RQ4: How scalable are varlen semantic IDs?}

\begin{table}[t]
\centering
\small
\setlength{\tabcolsep}{6pt}
\caption{
Scaling behavior of variable-length dVAE on Yambda.
$T$ denotes the maximum semantic ID length, and $|\mathcal{V}|$ denotes the vocabulary size.
}
\label{tab:rq4_scaling}
\begin{tabular}{cccccc}
\toprule
$\boldsymbol{T}$ & $\boldsymbol{|\mathcal{V}|}$
& $\boldsymbol{\mathcal{L}_{\text{recon}}}$
& $\boldsymbol{\mathcal{L}_{\text{recon}}^{\text{cold}}}$
& $\boldsymbol{\mathbb{E}_{p_{\text{data}}}[L]}$
& \textbf{PPL} \\
\midrule
5  & 4096   & 0.0254 & 0.0567 & 3.62 & 3575 \\
20 & 4096   & 0.0177 & 0.0300 & 8.33 & 2733 \\
5  & 32768  & 0.0150 & 0.0544 & 2.79 & 21350 \\
\bottomrule
\end{tabular}
\end{table}

We study how the proposed variable-length dVAE scales with respect to two key parameters: the maximum allowed semantic ID length and the vocabulary size. All experiments are conducted on Yambda under a fixed length penalty.

Table~\ref{tab:rq4_scaling} summarizes the results.
Increasing the maximum allowed length improves reconstruction quality, as the model can encode more information per item, but leads to longer semantic IDs. In contrast, increasing the vocabulary size yields simultaneous improvements in reconstruction quality and average code length: larger vocabularies allow the model to express finer distinctions using fewer symbols.

Importantly, the proposed method remains stable and continues to effectively utilize the available discrete capacity as these parameters increase.

Additional experiments are reported in Appendix~\ref{app:exp}.

\section{Conclusion}

We introduced variable-length semantic identifiers for recommender systems and proposed a principled training framework based on a discrete variational autoencoder with Gumbel-Softmax reparameterization.
Our method learns efficient item representations in which popular items receive shorter codes while rare and cold items are described with longer sequences, reflecting the non-uniform structure of real-world catalogs.
Empirically, we showed that variable-length semantic IDs preserve reconstruction quality, improve efficiency under fixed token budgets, and provide a stable alternative to REINFORCE-based approaches at scale.
These results position variable-length semantic IDs as a practical building block for large-scale generative retrieval and sequential recommendation models.

\begin{acks}
   This research was supported in part through computational resources of HPC facilities at HSE University\,\cite{hsecluster}.
\end{acks}
\newpage
\bibliographystyle{ACM-Reference-Format}
\bibliography{sample-base}

@String{Computing = "Computing" }

@String{Computer = "{IEEE} Computer" }

@String{Academic = "Academic Press" }

@String{Chelsea = "Chelsea" }

@String{Springer = "Springer-Verlag" }

@InProceedings{hstu,
  title = 	 {Actions Speak Louder than Words: Trillion-Parameter Sequential Transducers for Generative Recommendations},
  author =       {Zhai, Jiaqi and Liao, Lucy and Liu, Xing and Wang, Yueming and Li, Rui and Cao, Xuan and Gao, Leon and Gong, Zhaojie and Gu, Fangda and He, Jiayuan and Lu, Yinghai and Shi, Yu},
  booktitle = 	 {Proceedings of the 41st International Conference on Machine Learning},
  pages = 	 {58484--58509},
  year = 	 {2024},
  editor = 	 {Salakhutdinov, Ruslan and Kolter, Zico and Heller, Katherine and Weller, Adrian and Oliver, Nuria and Scarlett, Jonathan and Berkenkamp, Felix},
  volume = 	 {235},
  series = 	 {Proceedings of Machine Learning Research},
  month = 	 {21--27 Jul},
  publisher =    {PMLR},
  url = 	 {https://proceedings.mlr.press/v235/zhai24a.html}
}

@inproceedings{yambda,
author = {Ploshkin, Alexander and Tytskiy, Vladislav and Pismenny, Alexey and Baikalov, Vladimir and Taychinov, Evgeny and Permiakov, Artem and Burlakov, Daniil and Krofto, Eugene},
title = {Yambda-5B — A Large-Scale Multi-Modal Dataset for Ranking and Retrieval},
year = {2025},
isbn = {9798400713644},
publisher = {Association for Computing Machinery},
address = {New York, NY, USA},
url = {https://doi.org/10.1145/3705328.3748163},
doi = {10.1145/3705328.3748163},
booktitle = {Proceedings of the Nineteenth ACM Conference on Recommender Systems},
pages = {894–901},
numpages = {8},
location = {
},
series = {RecSys '25}
}

@inproceedings{logq,
author = {Yi, Xinyang and Yang, Ji and Hong, Lichan and Cheng, Derek Zhiyuan and Heldt, Lukasz and Kumthekar, Aditee and Zhao, Zhe and Wei, Li and Chi, Ed},
title = {Sampling-bias-corrected neural modeling for large corpus item recommendations},
year = {2019},
isbn = {9781450362436},
publisher = {Association for Computing Machinery},
address = {New York, NY, USA},
url = {https://doi.org/10.1145/3298689.3346996},
doi = {10.1145/3298689.3346996},
booktitle = {Proceedings of the 13th ACM Conference on Recommender Systems},
pages = {269–277},
numpages = {9},
location = {Copenhagen, Denmark},
series = {RecSys '19}
}

@inproceedings{correctlogq,
author = {Khrylchenko, Kirill and Baikalov, Vladimir and Makeev, Sergei and Matveev, Artem and Liamaev, Sergei},
title = {Correcting the LogQ Correction: Revisiting Sampled Softmax for Large-Scale Retrieval},
year = {2025},
isbn = {9798400713644},
publisher = {Association for Computing Machinery},
address = {New York, NY, USA},
url = {https://doi.org/10.1145/3705328.3748033},
doi = {10.1145/3705328.3748033},
booktitle = {Proceedings of the Nineteenth ACM Conference on Recommender Systems},
pages = {545–550},
numpages = {6},
location = {
},
series = {RecSys '25}
}

@inproceedings{p5,
author = {Geng, Shijie and Liu, Shuchang and Fu, Zuohui and Ge, Yingqiang and Zhang, Yongfeng},
title = {Recommendation as Language Processing (RLP): A Unified Pretrain, Personalized Prompt \& Predict Paradigm (P5)},
year = {2022},
isbn = {9781450392785},
publisher = {Association for Computing Machinery},
address = {New York, NY, USA},
url = {https://doi.org/10.1145/3523227.3546767},
doi = {10.1145/3523227.3546767},
booktitle = {Proceedings of the 16th ACM Conference on Recommender Systems},
pages = {299–315},
numpages = {17},
location = {Seattle, WA, USA},
series = {RecSys '22}
}

@misc{plum,
      title={PLUM: Adapting Pre-trained Language Models for Industrial-scale Generative Recommendations}, 
      author={Ruining He and Lukasz Heldt and Lichan Hong and Raghunandan Keshavan and Shifan Mao and Nikhil Mehta and Zhengyang Su and Alicia Tsai and Yueqi Wang and Shao-Chuan Wang and Xinyang Yi and Lexi Baugher and Baykal Cakici and Ed Chi and Cristos Goodrow and Ningren Han and He Ma and Romer Rosales and Abby Van Soest and Devansh Tandon and Su-Lin Wu and Weilong Yang and Yilin Zheng},
      year={2025},
      eprint={2510.07784},
      archivePrefix={arXiv},
      primaryClass={cs.IR},
      url={https://arxiv.org/abs/2510.07784}, 
}

@misc{onerec,
      title={OneRec: Unifying Retrieve and Rank with Generative Recommender and Iterative Preference Alignment}, 
      author={Jiaxin Deng and Shiyao Wang and Kuo Cai and Lejian Ren and Qigen Hu and Weifeng Ding and Qiang Luo and Guorui Zhou},
      year={2025},
      eprint={2502.18965},
      archivePrefix={arXiv},
      primaryClass={cs.IR},
      url={https://arxiv.org/abs/2502.18965}, 
}

@book{zla,
  author = {Zipf, George K.},
  publisher = {Addison-Wesley},
  title = {Human Behaviour and the Principle of Least Effort},
  year = 1949
}

@article{ec,
author = {Peters, Jannik and Waubert de Puiseau, Constantin and Tercan, Hasan and Gopikrishnan, Arya and Lucas de Carvalho, Gustavo Adolpho and Bitter, Christian and Meisen, Tobias},
title = {Emergent language: a survey and taxonomy},
year = {2025},
issue_date = {May 2025},
publisher = {Kluwer Academic Publishers},
address = {USA},
volume = {39},
number = {1},
issn = {1387-2532},
url = {https://doi.org/10.1007/s10458-025-09691-y},
doi = {10.1007/s10458-025-09691-y},
journal = {Autonomous Agents and Multi-Agent Systems},
month = mar,
numpages = {73}
}

@book{lewis,
  title={Convention: A Philosophical Study},
  author={Lewis, D.},
  isbn={9780470692967},
  url={https://books.google.ru/books?id=h1oBWq7JPCcC},
  year={2008},
  publisher={Wiley}
}

@inbook{anti-efficient,
author = {Chaabouni, Rahma and Kharitonov, Eugene and Dupoux, Emmanuel and Baroni, Marco},
title = {Anti-efficient encoding in emergent communication},
year = {2019},
publisher = {Curran Associates Inc.},
address = {Red Hook, NY, USA},
booktitle = {Proceedings of the 33rd International Conference on Neural Information Processing Systems},
articleno = {565},
numpages = {11}
}

@inproceedings{lazimpa,
    title = "``{L}az{I}mpa'': Lazy and Impatient neural agents learn to communicate efficiently",
    author = "Rita, Mathieu  and
      Chaabouni, Rahma  and
      Dupoux, Emmanuel",
    editor = "Fern{\'a}ndez, Raquel  and
      Linzen, Tal",
    booktitle = "Proceedings of the 24th Conference on Computational Natural Language Learning",
    month = nov,
    year = "2020",
    address = "Online",
    publisher = "Association for Computational Linguistics",
    url = "https://aclanthology.org/2020.conll-1.26/",
    doi = "10.18653/v1/2020.conll-1.26",
    pages = "335--343"
}

@inproceedings{ueda1,
    title = "On the Relationship between {Z}ipf{'}s Law of Abbreviation and Interfering Noise in Emergent Languages",
    author = "Ueda, Ryo  and
      Washio, Koki",
    editor = "Kabbara, Jad  and
      Lin, Haitao  and
      Paullada, Amandalynne  and
      Vamvas, Jannis",
    booktitle = "Proceedings of the 59th Annual Meeting of the Association for Computational Linguistics and the 11th International Joint Conference on Natural Language Processing: Student Research Workshop",
    month = aug,
    year = "2021",
    address = "Online",
    publisher = "Association for Computational Linguistics",
    url = "https://aclanthology.org/2021.acl-srw.6/",
    doi = "10.18653/v1/2021.acl-srw.6",
    pages = "60--70"
}

@inproceedings{ueda2,
 author = {Ueda, Ryo and Taniguchi, Tadahiro},
 booktitle = {International Conference on Learning Representations},
 editor = {B. Kim and Y. Yue and S. Chaudhuri and K. Fragkiadaki and M. Khan and Y. Sun},
 pages = {21729--21755},
 title = {Lewis\textquotesingle s Signaling Game as beta-VAE For Natural Word Lengths and Segments},
 url = {https://proceedings.iclr.cc/paper_files/paper/2024/file/5df4313ecd4875931fbdacc486cc1fcf-Paper-Conference.pdf},
 volume = {2024},
 year = {2024}
}

@misc{argus,
      title={Scaling Recommender Transformers to One Billion Parameters}, 
      author={Kirill Khrylchenko and Artem Matveev and Sergei Makeev and Vladimir Baikalov},
      year={2025},
      eprint={2507.15994},
      archivePrefix={arXiv},
      primaryClass={cs.IR},
      url={https://arxiv.org/abs/2507.15994}, 
}

@misc{vklsvd,
      title={VK-LSVD: A Large-Scale Industrial Dataset for Short-Video Recommendation}, 
      author={Aleksandr Poslavsky and Alexander D'yakonov and Yuriy Dorn and Andrey Zimovnov},
      year={2026},
      eprint={2602.04567},
      archivePrefix={arXiv},
      primaryClass={cs.IR},
      url={https://arxiv.org/abs/2602.04567}, 
}

@inproceedings{snapchat,
author = {Ju, Clark Mingxuan and Collins, Liam and Neves, Leonardo and Kumar, Bhuvesh and Wang, Louis Yufeng and Zhao, Tong and Shah, Neil},
title = {Generative Recommendation with Semantic IDs: A Practitioner's Handbook},
year = {2025},
isbn = {9798400720406},
publisher = {Association for Computing Machinery},
address = {New York, NY, USA},
url = {https://doi.org/10.1145/3746252.3761612},
doi = {10.1145/3746252.3761612},
booktitle = {Proceedings of the 34th ACM International Conference on Information and Knowledge Management},
pages = {6420–6425},
numpages = {6},
location = {Seoul, Republic of Korea},
series = {CIKM '25}
}

@inproceedings{time_to_split,
  title={Time to Split: Exploring Data Splitting Strategies for Offline Evaluation of Sequential Recommenders},
  author={Gusak, Danil and Volodkevich, Anna and Klenitskiy, Anton and Vasilev, Alexey and Frolov, Evgeny},
  booktitle={Proceedings of the 19th ACM Conference on Recommender Systems},
  doi={10.1145/3705328.3748164},
  year={2025}
}

@InProceedings{dalle1,
  title = 	 {Zero-Shot Text-to-Image Generation},
  author =       {Ramesh, Aditya and Pavlov, Mikhail and Goh, Gabriel and Gray, Scott and Voss, Chelsea and Radford, Alec and Chen, Mark and Sutskever, Ilya},
  booktitle = 	 {Proceedings of the 38th International Conference on Machine Learning},
  pages = 	 {8821--8831},
  year = 	 {2021},
  editor = 	 {Meila, Marina and Zhang, Tong},
  volume = 	 {139},
  series = 	 {Proceedings of Machine Learning Research},
  month = 	 {18--24 Jul},
  publisher =    {PMLR},
  url = 	 {https://proceedings.mlr.press/v139/ramesh21a.html}
}

@inproceedings{youtuberanking,
author = {Singh, Anima and Vu, Trung and Mehta, Nikhil and Keshavan, Raghunandan and Sathiamoorthy, Maheswaran and Zheng, Yilin and Hong, Lichan and Heldt, Lukasz and Wei, Li and Tandon, Devansh and Chi, Ed and Yi, Xinyang},
title = {Better Generalization with Semantic IDs: A Case Study in Ranking for Recommendations},
year = {2024},
isbn = {9798400705052},
publisher = {Association for Computing Machinery},
address = {New York, NY, USA},
url = {https://doi.org/10.1145/3640457.3688190},
doi = {10.1145/3640457.3688190},
booktitle = {Proceedings of the 18th ACM Conference on Recommender Systems},
pages = {1039–1044},
numpages = {6},
location = {Bari, Italy},
series = {RecSys '24}
}

@misc{amazondataset,
      title={Bridging Language and Items for Retrieval and Recommendation}, 
      author={Yupeng Hou and Jiacheng Li and Zhankui He and An Yan and Xiusi Chen and Julian McAuley},
      year={2024},
      eprint={2403.03952},
      archivePrefix={arXiv},
      primaryClass={cs.IR},
      url={https://arxiv.org/abs/2403.03952}, 
}

@inproceedings{relusq,
author = {So, David R. and Ma\'{n}ke, Wojciech and Liu, Hanxiao and Dai, Zihang and Shazeer, Noam and Le, Quoc V.},
title = {Primer: searching for efficient transformers for language modeling},
year = {2021},
isbn = {9781713845393},
publisher = {Curran Associates Inc.},
address = {Red Hook, NY, USA},
booktitle = {Proceedings of the 35th International Conference on Neural Information Processing Systems},
articleno = {460},
numpages = {13},
series = {NIPS '21}
}

@inproceedings{yarn,
  author       = {Weizhi Fei and
                  Xueyan Niu and
                  Pingyi Zhou and
                  Lu Hou and
                  Bo Bai and
                  Lei Deng and
                  Wei Han},
  editor       = {Lun{-}Wei Ku and
                  Andre Martins and
                  Vivek Srikumar},
  title        = {Extending Context Window of Large Language Models via Semantic Compression},
  booktitle    = {Findings of the Association for Computational Linguistics, {ACL} 2024,
                  Bangkok, Thailand and virtual meeting, August 11-16, 2024},
  series       = {Findings of {ACL}},
  volume       = {{ACL} 2024},
  pages        = {5169--5181},
  publisher    = {Association for Computational Linguistics},
  year         = {2024},
  url          = {https://doi.org/10.18653/v1/2024.findings-acl.306},
  doi          = {10.18653/V1/2024.FINDINGS-ACL.306},
  bibsource    = {dblp computer science bibliography, https://dblp.org}
}

@inproceedings{gs1,
  author       = {Eric Jang and
                  Shixiang Gu and
                  Ben Poole},
  title        = {Categorical Reparameterization with Gumbel-Softmax},
  booktitle    = {5th International Conference on Learning Representations, {ICLR} 2017,
                  Toulon, France, April 24-26, 2017, Conference Track Proceedings},
  publisher    = {OpenReview.net},
  year         = {2017},
  url          = {https://openreview.net/forum?id=rkE3y85ee},
  bibsource    = {dblp computer science bibliography, https://dblp.org}
}

@inproceedings{gs2,
  author       = {Chris J. Maddison and
                  Andriy Mnih and
                  Yee Whye Teh},
  title        = {The Concrete Distribution: {A} Continuous Relaxation of Discrete Random
                  Variables},
  booktitle    = {5th International Conference on Learning Representations, {ICLR} 2017,
                  Toulon, France, April 24-26, 2017, Conference Track Proceedings},
  publisher    = {OpenReview.net},
  year         = {2017},
  url          = {https://openreview.net/forum?id=S1jE5L5gl},
  bibsource    = {dblp computer science bibliography, https://dblp.org}
}

@article{reinforce,
author = {Williams, Ronald J.},
title = {Simple Statistical Gradient-Following Algorithms for Connectionist Reinforcement Learning},
year = {1992},
issue_date = {May 1992},
publisher = {Kluwer Academic Publishers},
address = {USA},
volume = {8},
number = {3–4},
issn = {0885-6125},
url = {https://doi.org/10.1007/BF00992696},
doi = {10.1007/BF00992696},
journal = {Mach. Learn.},
month = may,
pages = {229–256},
numpages = {28}
}

@inproceedings{shulman,
author = {Schulman, John and Heess, Nicolas and Weber, Theophane and Abbeel, Pieter},
title = {Gradient estimation using stochastic computation graphs},
year = {2015},
publisher = {MIT Press},
address = {Cambridge, MA, USA},
booktitle = {Proceedings of the 29th International Conference on Neural Information Processing Systems - Volume 2},
pages = {3528–3536},
numpages = {9},
location = {Montreal, Canada},
series = {NIPS'15}
}

@inproceedings{vq-vae,
author = {van den Oord, Aaron and Vinyals, Oriol and Kavukcuoglu, Koray},
title = {Neural discrete representation learning},
year = {2017},
isbn = {9781510860964},
publisher = {Curran Associates Inc.},
address = {Red Hook, NY, USA},
booktitle = {Proceedings of the 31st International Conference on Neural Information Processing Systems},
pages = {6309–6318},
numpages = {10},
location = {Long Beach, California, USA},
series = {NIPS'17}
}

@article{kmeans,
  author = {Hartigan, J. A. and Wong, M. A.},
  journal = {JSTOR: Applied Statistics},
  number = 1,
  pages = {100--108},
  title = {A k-means clustering algorithm},
  volume = 28,
  year = 1979
}

@inproceedings{dsi,
author = {Tay, Yi and Tran, Vinh Q. and Dehghani, Mostafa and Ni, Jianmo and Bahri, Dara and Mehta, Harsh and Qin, Zhen and Hui, Kai and Zhao, Zhe and Gupta, Jai and Schuster, Tal and Cohen, William W. and Metzler, Donald},
title = {Transformer memory as a differentiable search index},
year = {2022},
isbn = {9781713871088},
publisher = {Curran Associates Inc.},
address = {Red Hook, NY, USA},
booktitle = {Proceedings of the 36th International Conference on Neural Information Processing Systems},
articleno = {1587},
numpages = {13},
location = {New Orleans, LA, USA},
series = {NIPS '22}
}

@inproceedings{tiger,
author = {Rajput, Shashank and Mehta, Nikhil and Singh, Anima and Keshavan, Raghunandan and Vu, Trung and Heidt, Lukasz and Hong, Lichan and Tay, Yi and Tran, Vinh Q. and Samost, Jonah and Kula, Maciej and Chi, Ed H. and Sathiamoorthy, Maheswaran},
title = {Recommender systems with generative retrieval},
year = {2023},
publisher = {Curran Associates Inc.},
address = {Red Hook, NY, USA},
booktitle = {Proceedings of the 37th International Conference on Neural Information Processing Systems},
articleno = {452},
numpages = {17},
location = {New Orleans, LA, USA},
series = {NIPS '23}
}

@book{vq,
  title={Vector Quantization and Signal Compression},
  author={Gersho, A. and Gray, R.M.},
  isbn={9780792391814},
  lccn={lc91028580},
  series={The Springer International Series in Engineering and Computer Science},
  url={https://books.google.ru/books?id=DwcDm6xgItUC},
  year={1991},
  publisher={Springer US}
}

@article{ste,
  author       = {Yoshua Bengio and
                  Nicholas L{\'{e}}onard and
                  Aaron C. Courville},
  title        = {Estimating or Propagating Gradients Through Stochastic Neurons for
                  Conditional Computation},
  journal      = {CoRR},
  volume       = {abs/1308.3432},
  year         = {2013},
  url          = {http://arxiv.org/abs/1308.3432},
  eprinttype    = {arXiv},
  eprint       = {1308.3432},
  bibsource    = {dblp computer science bibliography, https://dblp.org}
}

@article{rq-vae,
author = {Zeghidour, Neil and Luebs, Alejandro and Omran, Ahmed and Skoglund, Jan and Tagliasacchi, Marco},
title = {SoundStream: An End-to-End Neural Audio Codec},
year = {2021},
issue_date = {2022},
publisher = {IEEE Press},
volume = {30},
issn = {2329-9290},
url = {https://doi.org/10.1109/TASLP.2021.3129994},
doi = {10.1109/TASLP.2021.3129994},
journal = {IEEE/ACM Trans. Audio, Speech and Lang. Proc.},
month = nov,
pages = {495–507},
numpages = {13}
}

@inproceedings{vae,
  author = {Kingma, Diederik P. and Welling, Max},
  booktitle = {2nd International Conference on Learning Representations, {ICLR} 2014, Banff, AB, Canada, April 14-16, 2014, Conference Track Proceedings},
  eprint = {http://arxiv.org/abs/1312.6114v10},
  eprintclass = {stat.ML},
  eprinttype = {arXiv},
  title = {{Auto-Encoding Variational Bayes}},
  year = 2014
}

@inproceedings{freebits,
author = {Kingma, Diederik P. and Salimans, Tim and Jozefowicz, Rafal and Chen, Xi and Sutskever, Ilya and Welling, Max},
title = {Improved variational inference with inverse autoregressive flow},
year = {2016},
isbn = {9781510838819},
publisher = {Curran Associates Inc.},
address = {Red Hook, NY, USA},
booktitle = {Proceedings of the 30th International Conference on Neural Information Processing Systems},
pages = {4743–4751},
numpages = {9},
location = {Barcelona, Spain},
series = {NIPS'16}
}

@inproceedings{betavae,
  author       = {Irina Higgins and
                  Lo{\"{\i}}c Matthey and
                  Arka Pal and
                  Christopher P. Burgess and
                  Xavier Glorot and
                  Matthew M. Botvinick and
                  Shakir Mohamed and
                  Alexander Lerchner},
  title        = {beta-VAE: Learning Basic Visual Concepts with a Constrained Variational
                  Framework},
  booktitle    = {5th International Conference on Learning Representations, {ICLR} 2017,
                  Toulon, France, April 24-26, 2017, Conference Track Proceedings},
  publisher    = {OpenReview.net},
  year         = {2017},
  url          = {https://openreview.net/forum?id=Sy2fzU9gl},
  bibsource    = {dblp computer science bibliography, https://dblp.org}
}

@inproceedings{titov,
author = {Havrylov, Serhii and Titov, Ivan},
title = {Emergence of language with multi-agent games: learning to communicate with sequences of symbols},
year = {2017},
isbn = {9781510860964},
publisher = {Curran Associates Inc.},
address = {Red Hook, NY, USA},
booktitle = {Proceedings of the 31st International Conference on Neural Information Processing Systems},
pages = {2146–2156},
numpages = {11},
location = {Long Beach, California, USA},
series = {NIPS'17}
}

@inproceedings{rkmeans,
author = {Luo, Xinchen and Cao, Jiangxia and Sun, Tianyu and Yu, Jinkai and Huang, Rui and Yuan, Wei and Lin, Hezheng and Zheng, Yichen and Wang, Shiyao and Hu, Qigen and Qiu, Changqing and Zhang, Jiaqi and Zhang, Xu and Yan, Zhiheng and Zhang, Jingming and Zhang, Simin and Wen, Mingxing and Liu, Zhaojie and Zhou, Guorui},
title = {QARM: Quantitative Alignment Multi-Modal Recommendation at Kuaishou},
year = {2025},
isbn = {9798400720406},
publisher = {Association for Computing Machinery},
address = {New York, NY, USA},
url = {https://doi.org/10.1145/3746252.3761502},
doi = {10.1145/3746252.3761502},
booktitle = {Proceedings of the 34th ACM International Conference on Information and Knowledge Management},
pages = {5915–5922},
numpages = {8},
location = {Seoul, Republic of Korea},
series = {CIKM '25}
}

@misc{nanochat,
  author = {Andrej Karpathy},
  title = {nanochat: The best ChatGPT that \$100 can buy},
  year = {2025},
  publisher = {GitHub},
  url = {https://github.com/karpathy/nanochat}
}

@misc{nanogpt_speedrun,
  author       = {Keller Jordan and Jeremy Bernstein and Brendan Rappazzo and
                  @fernbear.bsky.social and Boza Vlado and You Jiacheng and
                  Franz Cesista and Braden Koszarsky and @Grad62304977},
  title        = {modded-nanogpt: Speedrunning the NanoGPT baseline},
  year         = {2024},
  url          = {https://github.com/KellerJordan/modded-nanogpt}
}

@misc{muon,
  author       = {Keller Jordan and Yuchen Jin and Vlado Boza and Jiacheng You and
                  Franz Cesista and Laker Newhouse and Jeremy Bernstein},
  title        = {Muon: An optimizer for hidden layers in neural networks},
  year         = {2024},
  url          = {https://kellerjordan.github.io/posts/muon/}
}

@inproceedings{
lazaridou,
title={Multi-Agent Cooperation and the Emergence of (Natural) Language},
author={Angeliki Lazaridou and Alexander Peysakhovich and Marco Baroni},
booktitle={International Conference on Learning Representations},
year={2017},
url={https://openreview.net/forum?id=Hk8N3Sclg}
}

@inproceedings{music_semantics,
author = {Mei, M. Jeffrey and Henkel, Florian and Sandberg, Samuel E. and Bembom, Oliver and Ehmann, Andreas F.},
title = {Semantic IDs for Music Recommendation},
year = {2025},
isbn = {9798400713644},
publisher = {Association for Computing Machinery},
address = {New York, NY, USA},
url = {https://doi.org/10.1145/3705328.3748139},
doi = {10.1145/3705328.3748139},
booktitle = {Proceedings of the Nineteenth ACM Conference on Recommender Systems},
pages = {1070–1073},
numpages = {4},
location = {
},
series = {RecSys '25}
}

@article{linkedin_semantic,
  author       = {Fedor Borisyuk and
                  Lars Hertel and
                  Ganesh Parameswaran and
                  Gaurav Srivastava and
                  Sudarshan Srinivasa Ramanujam and
                  Borja Ocejo and
                  Peng Du and
                  Andrei Akterskii and
                  Neil Daftary and
                  Shao Tang and
                  Daqi Sun and
                  Qiang Charles Xiao and
                  Deepesh Nathani and
                  Mohit Kothari and
                  Yun Dai and
                  Aman Gupta},
  title        = {From Features to Transformers: Redefining Ranking for Scalable Impact},
  journal      = {CoRR},
  volume       = {abs/2502.03417},
  year         = {2025},
  url          = {https://doi.org/10.48550/arXiv.2502.03417},
  doi          = {10.48550/ARXIV.2502.03417},
  eprinttype    = {arXiv},
  eprint       = {2502.03417},
  bibsource    = {dblp computer science bibliography, https://dblp.org}
}

@inproceedings{spotify_semantic,
author = {Penha, Gustavo and D'Amico, Edoardo and De Nadai, Marco and Palumbo, Enrico and Tamborrino, Alexandre and Vardasbi, Ali and Lefarov, Max and Lin, Shawn and Heath, Timothy and Fabbri, Francesco and Bouchard, Hugues},
title = {Semantic IDs for Joint Generative Search and Recommendation},
year = {2025},
isbn = {9798400713644},
publisher = {Association for Computing Machinery},
address = {New York, NY, USA},
url = {https://doi.org/10.1145/3705328.3759300},
doi = {10.1145/3705328.3759300},
booktitle = {Proceedings of the Nineteenth ACM Conference on Recommender Systems},
pages = {1296–1301},
numpages = {6},
location = {
},
series = {RecSys '25}
}

@inproceedings{egg,
    title = "{EGG}: a toolkit for research on Emergence of lan{G}uage in Games",
    author = "Kharitonov, Eugene  and
      Chaabouni, Rahma  and
      Bouchacourt, Diane  and
      Baroni, Marco",
    editor = "Pad{\'o}, Sebastian  and
      Huang, Ruihong",
    booktitle = "Proceedings of the 2019 Conference on Empirical Methods in Natural Language Processing and the 9th International Joint Conference on Natural Language Processing (EMNLP-IJCNLP): System Demonstrations",
    month = nov,
    year = "2019",
    address = "Hong Kong, China",
    publisher = "Association for Computational Linguistics",
    url = "https://aclanthology.org/D19-3010/",
    doi = "10.18653/v1/D19-3010",
    pages = "55--60"
}

@misc{gemma,
      title={EmbeddingGemma: Powerful and Lightweight Text Representations}, 
      author={Henrique Schechter Vera and Sahil Dua and Biao Zhang and Daniel Salz and Ryan Mullins and Sindhu Raghuram Panyam and Sara Smoot and Iftekhar Naim and Joe Zou and Feiyang Chen and Daniel Cer and Alice Lisak and Min Choi and Lucas Gonzalez and Omar Sanseviero and Glenn Cameron and Ian Ballantyne and Kat Black and Kaifeng Chen and Weiyi Wang and Zhe Li and Gus Martins and Jinhyuk Lee and Mark Sherwood and Juyeong Ji and Renjie Wu and Jingxiao Zheng and Jyotinder Singh and Abheesht Sharma and Divyashree Sreepathihalli and Aashi Jain and Adham Elarabawy and AJ Co and Andreas Doumanoglou and Babak Samari and Ben Hora and Brian Potetz and Dahun Kim and Enrique Alfonseca and Fedor Moiseev and Feng Han and Frank Palma Gomez and Gustavo Hernández Ábrego and Hesen Zhang and Hui Hui and Jay Han and Karan Gill and Ke Chen and Koert Chen and Madhuri Shanbhogue and Michael Boratko and Paul Suganthan and Sai Meher Karthik Duddu and Sandeep Mariserla and Setareh Ariafar and Shanfeng Zhang and Shijie Zhang and Simon Baumgartner and Sonam Goenka and Steve Qiu and Tanmaya Dabral and Trevor Walker and Vikram Rao and Waleed Khawaja and Wenlei Zhou and Xiaoqi Ren and Ye Xia and Yichang Chen and Yi-Ting Chen and Zhe Dong and Zhongli Ding and Francesco Visin and Gaël Liu and Jiageng Zhang and Kathleen Kenealy and Michelle Casbon and Ravin Kumar and Thomas Mesnard and Zach Gleicher and Cormac Brick and Olivier Lacombe and Adam Roberts and Qin Yin and Yunhsuan Sung and Raphael Hoffmann and Tris Warkentin and Armand Joulin and Tom Duerig and Mojtaba Seyedhosseini},
      year={2025},
      eprint={2509.20354},
      archivePrefix={arXiv},
      primaryClass={cs.CL},
      url={https://arxiv.org/abs/2509.20354}, 
}

@article{normuon,
  author       = {Zichong Li and
                  Liming Liu and
                  Chen Liang and
                  Weizhu Chen and
                  Tuo Zhao},
  title        = {NorMuon: Making Muon more efficient and scalable},
  journal      = {CoRR},
  volume       = {abs/2510.05491},
  year         = {2025},
  url          = {https://doi.org/10.48550/arXiv.2510.05491},
  doi          = {10.48550/ARXIV.2510.05491},
  eprinttype    = {arXiv},
  eprint       = {2510.05491},
  bibsource    = {dblp computer science bibliography, https://dblp.org}
}

@inbook{rmsnorm,
author = {Zhang, Biao and Sennrich, Rico},
title = {Root mean square layer normalization},
year = {2019},
publisher = {Curran Associates Inc.},
address = {Red Hook, NY, USA},
booktitle = {Proceedings of the 33rd International Conference on Neural Information Processing Systems},
articleno = {1110},
numpages = {12}
}

@inproceedings{transformer,
author = {Vaswani, Ashish and Shazeer, Noam and Parmar, Niki and Uszkoreit, Jakob and Jones, Llion and Gomez, Aidan N. and Kaiser, \L{}ukasz and Polosukhin, Illia},
title = {Attention is all you need},
year = {2017},
isbn = {9781510860964},
publisher = {Curran Associates Inc.},
address = {Red Hook, NY, USA},
booktitle = {Proceedings of the 31st International Conference on Neural Information Processing Systems},
pages = {6000–6010},
numpages = {11},
location = {Long Beach, California, USA},
series = {NIPS'17}
}

@misc{gemini,
  title={Gemini: A Family of Highly Capable Multimodal Models}, 
  author={Gemini Team},
  year={2024},
  eprint={2312.11805},
  archivePrefix={arXiv},
  primaryClass={cs.CL},
  url={https://arxiv.org/abs/2312.11805}, 
}

@article{hsecluster,
doi = {10.1088/1742-6596/1740/1/012050},
url = {https://doi.org/10.1088/1742-6596/1740/1/012050},
year = {2021},
month = {jan},
publisher = {IOP Publishing},
volume = {1740},
number = {1},
pages = {012050},
author = {Kostenetskiy, P. S. and Chulkevich, R. A. and Kozyrev, V. I.},
title = {HPC Resources of the Higher School of Economics},
journal = {Journal of Physics: Conference Series}
}

\pagebreak
\appendix
\onecolumn

\appendix
\onecolumn

\newpage
\section{Detailed ELBO Derivation}
\label{app:elbo_derivation}

We derive the evidence lower bound (ELBO) for the proposed variable-length latent representation.

\subsection{Model Definition}

\paragraph{Generative model.}
Let $L \in \{1, \dots, T\}$ be a discrete latent length variable and
$z = (z_1, \dots, z_T)$ a sequence of latent tokens, where $z_t \in \mathcal V \cup \{\texttt{pad}\}$.
The joint generative model factorizes as
\begin{equation}
p(x, z, L) = p(L)\, p(z \mid L)\, p(x \mid z, L),
\end{equation}
where the conditional prior over $z$ is defined with padding as
\begin{equation}
p(z \mid L)
=
\prod_{t=1}^{L} p_{\mathcal V}(z_t)
\prod_{t=L+1}^{T} \mathbf 1\{z_t = \texttt{pad}\}.
\end{equation}
The likelihood depends only on the prefix $z_{\le L}$:
\begin{equation}
p(x \mid z, L) = p(x \mid z_{1:L}, L).
\end{equation}

\paragraph{Length prior.}
We use a truncated geometric distribution as a length prior:
\begin{equation}
p(L=l)=\frac{(1-\alpha)^{l-1}\alpha}{1-(1-\alpha)^T},
\qquad l \in \{1,\dots,T\},
\end{equation}
where $\alpha \in (0,1)$ is the per-step stopping probability.

\paragraph{Approximate posterior.}
We define the variational distribution as
\begin{equation}
q(z, L \mid x) =
\left(\prod_{t=1}^{L} q(z_t \mid z_{1:t-1}, x)\right)\;
q(L \mid x, z_{1:L})\;
\left(\prod_{t=L+1}^{T} \mathbb{I}\{z_t = \texttt{pad}\}\right).
\end{equation}
By construction, $q$ has the same support as $p(z\mid L)$: tokens after the stopping time are deterministically equal to \texttt{pad}.
This avoids undefined terms of the form $\log \mathbf 1\{z_t=\texttt{pad}\}$.

\subsection{Derivation}

The ELBO is given by
\begin{equation}
\mathcal L
=
\mathbb{E}_{q(z, L \mid x)}\!\left[ \log p(x \mid z_{1:L}) \right]
-
\mathrm{KL}\!\left( q(z, L \mid x)\,\|\, p(z, L) \right).
\end{equation}

\subsubsection{Reconstruction term}

We start with the reconstruction term and marginalize over the latent length:
\begin{align}
\mathbb{E}_{q(z, L \mid x)}\!\left[ \log p(x \mid z_{1:L}) \right]
&=
\mathbb{E}_{q(z \mid x)} \mathbb{E}_{q(L \mid x, z)} \!\left[ \log p(x \mid z_{1:L}) \right] =
\mathbb{E}_{q(z \mid x)} \!\left[ \sum_{l=1}^T q(L = l \mid x, z_{1:l}) \log p(x \mid z_{1:l}) \right]. \label{eq:recon_sum}
\end{align}
In practice, this corresponds to aggregating reconstruction errors over all prefixes, weighted by their probabilities.

\subsubsection{KL decomposition}

Next, we decompose the KL divergence:
\begin{align}
\mathrm{KL}\!\left( q(z, L \mid x)\,\|\, p(z, L) \right)
&=
\mathbb{E}_{q(z, L \mid x)}\!\left[ \log q(z, L \mid x) - \log p(z, L) \right]. \label{eq:kl_def}
\end{align}
Using the definitions of $q$ and $p$,
\begin{align}
\log q(z,L\mid x)
&=
\sum_{t=1}^{L} \log q(z_t \mid x, z_{1:t-1})
+ \log q(L \mid x, z_{1:L})
+ \sum_{t=L+1}^{T} \log \mathbf 1\{z_t=\texttt{pad}\}, \\
\log p(z,L)
&=
\log p(L)
+ \sum_{t=1}^{L} \log p_{\mathcal V}(z_t)
+ \sum_{t=L+1}^{T} \log \mathbf 1\{z_t=\texttt{pad}\}.
\end{align}
The indicator terms cancel on the support of $q$, hence
\begin{align}
\mathrm{KL}\!\left( q(z, L \mid x)\,\|\, p(z, L) \right)
&=
\underbrace{\mathbb{E}_{q(z, L \mid x)}\!\left[ \log q(L \mid x, z_{1:L}) - \log p(L) \right]}_{\text{length regularization}} +\underbrace{\mathbb{E}_{q(z, L \mid x)}\!\left[ \sum_{t=1}^{L} \left( \log q(z_t \mid x, z_{1:t-1}) - \log p_{\mathcal V}(z_t) \right) \right]}_{\text{vocabulary regularization}}.
\label{eq:kl_split}
\end{align}

\subsubsection{Length regularization}

Consider the first term in~\eqref{eq:kl_split}:
\begin{align}
\mathbb{E}_{q(z, L \mid x)}\!\left[ \log q(L \mid x, z_{1:L}) - \log p(L) \right]
&=
\mathbb{E}_{q(z \mid x)} \mathbb{E}_{q(L \mid x, z)}\!\left[ \log q(L \mid x, z) - \log p(L) \right] \nonumber\\
&=
\mathbb{E}_{q(z \mid x)}\!\left[ -H\!\left(q(L \mid x, z)\right) \right]
-\mathbb{E}_{q(z \mid x)} \mathbb{E}_{q(L \mid x, z)}\!\left[\log p(L)\right]. \label{eq:length_term_start}
\end{align}

Substituting the truncated geometric prior,
\begin{align}
\mathbb{E}_{q(L \mid x, z)}[\log p(L)]
&=
\sum_{l=1}^{T} q(L=l \mid x, z)\,
\log\left(\frac{(1-\alpha)^{l-1}\alpha}{1-(1-\alpha)^T}\right) =
\sum_{l=1}^{T} q(L=l \mid x, z)\,
\Big( (l-1)\log(1-\alpha) + \log\alpha - \log(1-(1-\alpha)^T) \Big).
\end{align}

The last two terms are constant w.r.t.\ $l$, hence
\begin{align}
\mathbb{E}_{q(L \mid x, z)}[\log p(L)]
&=
\underbrace{\log(1-\alpha)}_{-\lambda}\, \mathbb{E}_{q(L \mid x, z)}[L] + \text{Const}. \label{eq:geom_expectation}
\end{align}

Combining~\eqref{eq:length_term_start} and~\eqref{eq:geom_expectation}, we obtain (up to an additive constant):
\begin{align}
\mathbb{E}_{q(z, L \mid x)}\!\left[ \log q(L \mid x, z) - \log p(L) \right]
&=
\mathbb{E}_{q(z \mid x)}\!\left[ -H\!\left(q(L \mid x, z)\right) \right]
+\lambda\,\mathbb{E}_{q(z \mid x)}\mathbb{E}_{q(L \mid x, z)}[L]
+ \text{Const}. \label{eq:length_final}
\end{align}
This yields the length-regularization term consisting of an expected-length penalty and an entropy term.

\subsubsection{Vocabulary regularization}
Now consider the second term in~\eqref{eq:kl_split}:
\begin{align}
&\mathbb{E}_{q(z, L \mid x)}\!\left[ \sum_{t=1}^{L}
\left( \log q(z_t \mid x, z_{1:t-1}) - \log p_{\mathcal V}(z_t) \right) \right] \nonumber\\
&\qquad=
\mathbb{E}_{q(z \mid x)}\!
\mathbb{E}_{q(L \mid x, z)}\!\left[
\sum_{t=1}^{T} \mathbf 1\{L \ge t\}
\left( \log q(z_t \mid x, z_{1:t-1}) - \log p_{\mathcal V}(z_t) \right)
\right] \nonumber\\
&\qquad=
\mathbb{E}_{q(z \mid x)}\!\left[
\sum_{t=1}^{T}
\mathbb{E}_{q(L \mid x, z)}\!\left[\mathbf 1\{L \ge t\}\right]
\mathbb{E}_{q(z_t \mid x, z_{1:t-1})}\!\left[
\log q(z_t \mid x, z_{1:t-1}) - \log p_{\mathcal V}(z_t)
\right]
\right] \nonumber\\
&\qquad=
\mathbb{E}_{q(z \mid x)}\!\left[
\sum_{t=1}^{T}
\sum_{l=t}^{T} q(L=l \mid x, z_{1:t})\;
\mathbb{E}_{q(z_t \mid x, z_{1:t-1})}\!\left[
\log q(z_t \mid x, z_{1:t-1}) - \log p_{\mathcal V}(z_t)
\right]
\right] \nonumber\\
&\qquad=
\mathbb{E}_{q(z \mid x)}\!\left[
\sum_{t=1}^{T}
q(L \ge t \mid x, z_{1:t})\;
\mathbb{E}_{q(z_t \mid x, z_{1:t-1})}\!\left[
\log q(z_t \mid x, z_{1:t-1}) - \log p_{\mathcal V}(z_t)
\right]
\right] \nonumber\\
&\qquad=
\mathbb{E}_{q(z \mid x)}\!\left[
\sum_{t=1}^{T}
q(L \ge t \mid x, z_{1:t})\;
\mathrm{KL}\!\left(
q(z_t \mid x, z_{1:t-1}) \,\Vert\, p_{\mathcal V}
\right)
\right],
\label{eq:vocab_alive_kl}
\end{align}
where we use the convention $q(z_1 \mid x, z_{1:0}) = q(z_1 \mid x)$.

\begin{algorithm}[H]
\caption{Training objective computation for variable-length dVAE}
\label{alg:varlen_dvae_loss}
\begin{algorithmic}[1]
\Require embedding $x\in\mathbb{R}^d$, max length $T$, vocab size $V$, temperature $\tau$
\Require length cost $\lambda$, optional weight $\beta$, optional free-bits threshold $\delta$
\Statex
\Function{ForwardLoss}{$x,\tau,\lambda,\beta,\delta$}
    \State \textbf{/* Encoder: sample relaxed message up to maxlen via Gumbel-Softmax */}
    \State $h \gets \mathrm{Backbone}(x)$
    \For{$t=1$ \textbf{to} $T$}
        \State $\ell_t \gets \mathrm{TokenLogits}(h, t)$ \Comment{$\ell_t\in\mathbb{R}^V$}
        \State $g \sim \mathrm{Gumbel}(0,1)^V$
        \State $m_t \gets \mathrm{softmax}\big((\ell_t + g)/\tau\big)$ \Comment{relaxed one-hot, $m_t\in\Delta^{V-1}$}
        \State Store $\ell_t$ and $m_t$
        \If{$t < T$}
            \State $b_t \gets m_t^\top C_t$ \Comment{expected codebook vector}
            \State $h \gets \mathrm{RMSNorm}\big(h - \exp(\gamma_t)\, b_t\big)$ \Comment{residual update}
            \State Store $h$ as a stop-feature for step $t{+}1$
        \EndIf
    \EndFor
    \Statex

    \State \textbf{/* Encoder: length distribution (no Gumbel-Softmax) */}
    \State $\eta \gets \mathrm{LengthLogits}(\text{stop-features})$ \Comment{$\eta\in\mathbb{R}^T$}
    \State $q_L \gets \mathrm{softmax}(\eta)$ \Comment{$q_L[t]=q(L=t)$}
    \State $a \gets \mathrm{AliveFromLength}(q_L)$ \Comment{$a[t]=\sum_{k=t}^{T} q_L[k]$}
    \Statex

    \State \textbf{/* Decoder: causal, returns reconstructions for all prefixes in one pass */}
    \State $(\hat{x}_1,\dots,\hat{x}_T) \gets \mathrm{Decoder}(m_1,\dots,m_T)$
    \Statex

    \State \textbf{/* Reconstruction loss: weighted sum over prefix reconstructions */}
    \State $\tilde{q}_L \gets 0.9\cdot q_L + 0.1\cdot (1/T)$ \Comment{smoothing used only here}
    \State $\mathcal{L}_{\mathrm{recon}} \gets 0$
    \For{$t=1$ \textbf{to} $T$}
        \State $\ell^{\mathrm{rec}}_t \gets \mathrm{MSE}(x,\hat{x}_t)$ \Comment{sum over embedding dims}
        \State $\mathcal{L}_{\mathrm{recon}} \gets \mathcal{L}_{\mathrm{recon}} + \tilde{q}_L[t]\cdot \ell^{\mathrm{rec}}_t$
    \EndFor
    \Statex

    \State \textbf{/* Vocabulary regularization: per-step KL to uniform, alive-weighted */}
    \State $\mathcal{L}_{\mathrm{reg}}^{\mathrm{vocab}} \gets 0$
    \For{$t=1$ \textbf{to} $T$}
        \State $q_t \gets \mathrm{softmax}(\ell_t)$
        \State $H_t \gets -\sum_{i=1}^{V} q_t[i]\log q_t[i]$
        \State $\mathrm{KL}_t \gets \log V - H_t$ \Comment{$\mathrm{KL}(q_t \Vert \mathrm{Unif})$}
        \If{$\delta$ is provided} 
            \State $\mathrm{KL}_t \gets \max(0,\mathrm{KL}_t-\delta)$ \Comment{free-bits}
        \EndIf
        \State $\mathcal{L}_{\mathrm{reg}}^{\mathrm{vocab}} \gets \mathcal{L}_{\mathrm{reg}}^{\mathrm{vocab}} + a[t]\cdot \mathrm{KL}_t$
    \EndFor
    \Statex

    \State \textbf{/* Length regularization: expected length + entropy of length posterior */}
    \State $E_L \gets \sum_{t=1}^{T} t\cdot q_L[t]$
    \State $H_L \gets -\sum_{t=1}^{T} q_L[t]\log q_L[t]$
    \State $\mathcal{L}_{\mathrm{reg}}^{\mathrm{length}} \gets \lambda\cdot E_L - H_L$
    \Statex

    \State \textbf{/* Total loss */}
    \If{$\beta$ is provided}
        \State $\mathcal{L} \gets \mathcal{L}_{\mathrm{recon}} + \beta\cdot\big(\mathcal{L}_{\mathrm{reg}}^{\mathrm{vocab}}+\mathcal{L}_{\mathrm{reg}}^{\mathrm{length}}\big)$
    \Else
        \State $\mathcal{L} \gets \mathcal{L}_{\mathrm{recon}}$
    \EndIf
    \State \Return $\mathcal{L}$
\EndFunction
\end{algorithmic}
\end{algorithm}

\newpage
\section{Additional Experiments}\label{app:exp}
In the appendix, we provide additional diagnostics characterizing how information is distributed across positions in semantic IDs.

\paragraph{Progressive Reconstruction}

\begin{table}[t]
\centering
\small
\setlength{\tabcolsep}{6pt}
\caption{Reconstruction loss from semantic ID prefixes on Yambda}
\label{tab:prefix_recon}
\begin{tabular}{lccccc}
\toprule
\textbf{Method} & $\mathcal{L}_{\text{recon}}$@1 & $\mathcal{L}_{\text{recon}}$@2 & $\mathcal{L}_{\text{recon}}$@3 & $\mathcal{L}_{\text{recon}}$@4 & $\mathcal{L}_{\text{recon}}$@5 \\
\midrule
R-KMeans & 
0.1644 & 0.0959 & 0.0665 & 0.0494 & 0.0384 \\
dVAE (fixed-length) & 
0.1400 & 0.0678 & 0.0512 & 0.0355 & 0.0246 \\
dVAE (varlen) & 
0.1344 & 0.0606 & 0.0404 & 0.0313 & 0.0270 \\
\bottomrule
\end{tabular}

\vspace{0.3em}
\small
\end{table}

Table~\ref{tab:prefix_recon} reports reconstruction quality obtained from intermediate semantic ID prefixes.
For R-KMeans, this behavior is expected by construction, as residual quantization progressively refines the representation by subtracting selected centroids at each step.

Interestingly, the fixed-length dVAE also exhibits meaningful reconstruction from intermediate prefixes, despite being trained using reconstruction loss applied only to the final prefix.

The variable-length dVAE further strengthens this effect.
By explicitly optimizing reconstruction quality across prefixes and learning when to stop, the varlen model achieves consistently better reconstruction from short prefixes compared to its fixed-length counterpart.

\paragraph{Position-wise Vocabulary Utilization}

\begin{table}[t]
\centering
\small
\setlength{\tabcolsep}{6pt}
\caption{Position-wise token perplexity on Yambda (higher means more diverse token usage)}
\label{tab:pos_ppl}
\begin{tabular}{lccccccc}
\toprule
\textbf{Method} & PPL@1 & PPL@2 & PPL@3 & PPL@4 & PPL@5 \\
\midrule
R-KMeans &
3895 & 3674 & 3489 & 3414 & 3393  \\
dVAE (fixed-length) &
1959 & 3366 & 3315 & 3284 & 3135  \\
dVAE (varlen) &
2168 & 3383 & 3215 & 3167 & 3175  \\
REINFORCE (fixed-length) &
1340 & 2146 & 2689 & 3035 & 3232  \\
\bottomrule
\end{tabular}
\end{table}

Table~\ref{tab:pos_ppl} reports token perplexity separately for each position in the semantic ID; for variable-length models, perplexity at position $t$ is computed only over items whose generated length includes this position (i.e., $L \ge t$).

R-KMeans exhibits substantially higher perplexity at the first position, indicating that hard residual quantization spreads information aggressively into the initial token.
In contrast, dVAE-based models use a less diverse set of tokens at the first position while maintaining high perplexity at later positions.

\paragraph{Position-wise Vocabulary Utilization for Long Codes}

\begin{figure}[t]
    \centering
    \includegraphics[width=0.6\linewidth]{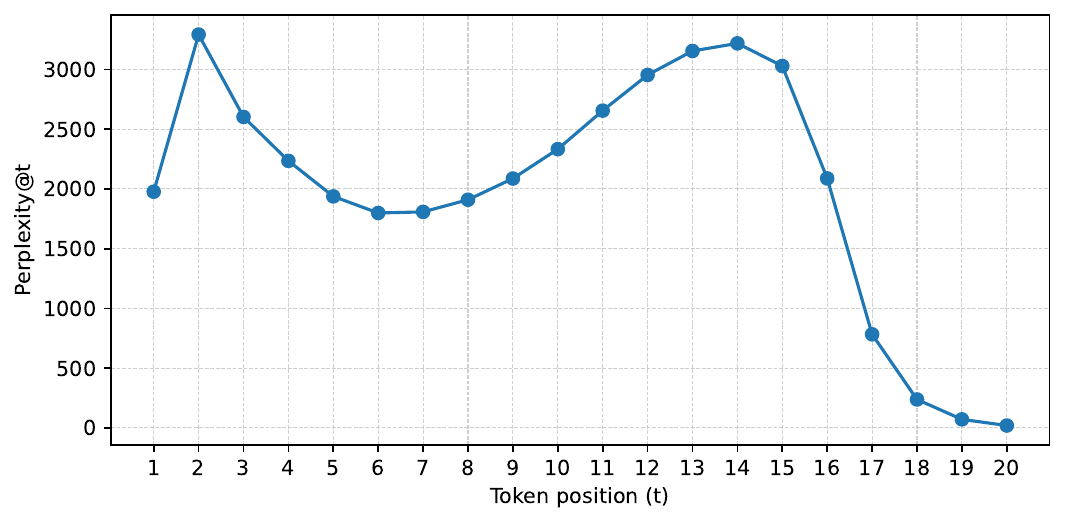}
    \caption{
    Position-wise token perplexity for varlen dVAE with $\texttt{maxlen}=20$ on Yambda
    }
    \label{fig:perplexity_by_position}
\end{figure}

To further analyze how the effective vocabulary usage evolves along the code,
we report position-wise token perplexity for the variable-length dVAE model
with $\texttt{maxlen}=20$.

Figure~\ref{fig:perplexity_by_position} shows that token perplexity
initially increases and reaches its maximum around the 10--15-th positions.
This indicates that the model progressively utilizes a richer subset of the vocabulary
when encoding finer-grained semantic distinctions.
Beyond this point, token perplexity rapidly decreases by more than two orders of magnitude,
suggesting that later positions employ a highly restricted set of tokens.

This sharp drop implies that long suffixes rely on a very small effective vocabulary
and are emitted only for a small subset of items.

\end{document}